\documentclass[prd,twocolumn,showpacs,preprintnumbers,nofootinbib]{revtex4}
\usepackage{amsfonts,amsmath,amssymb}
\usepackage{graphicx}
\usepackage{color}
\usepackage[plainpages=false, colorlinks=true, anchorcolor=blue, linkcolor=blue, citecolor=blue, bookmarks=false]{hyperref}
\usepackage{natbib}
\pdfoutput=1
\usepackage{natbib}
\usepackage{breqn}

\makeatletter
\let\cat@comma@active\@empty
\makeatother

\begin{document}
%\markboth{Authors' Names}

%%%%%%%% Journals %%%%%%%%%%%%%%
\newcommand{\apjl}{Astrophys. J. Lett.}
\newcommand{\apjs}{Astrophys. J. Suppl. Ser.}
\newcommand{\aap}{Astron. \& Astrophys.}
\newcommand{\aj}{Astron. J.}
\newcommand{\araa}{Ann. Rev. Astron. Astrophys. } %ARA$\&$A}
\newcommand{\mnras}{Mon. Not. R. Astron. Soc.}
\newcommand{\apss}{Astrophysics \& Space Sciences}
\newcommand{\jcap}{JCAP}
\newcommand{\pasj}{PASJ}
\newcommand{\pasp}{PASP}
\newcommand{\pasa}{Pub. Astro. Soc. Aust.}
\newcommand{\physrep}{Phys. Rep.}

%%%%%%%%%%%%%%%%%% TITLE %%%%%%%%%%%%%%%%%%%%%%%%%%%%%%%%%%%% 
\title{Limit on graviton mass using stacked galaxy cluster catalogs from SPT-SZ, Planck-SZ and SDSS-redMaPPer}
\author{Sajal \surname{Gupta}$^1$}
\altaffiliation{E-mail: sg15ms084@iiserkol.ac.in}
\author{Shantanu  \surname{Desai}$^2$} \altaffiliation{E-mail: shntn05@gmail.com}

\affiliation{$^{1}$Department of Physical Sciences, IISER-Kolkata, Mohanpur, West Bengal-741246, India}

\affiliation{$^{2}$Department of Physics, Indian Institute of Technology, Hyderabad, Telangana-502285, India}

\begin{abstract}
In the last few years, there has been a resurgence of interest in obtaining observational bounds  on the graviton mass,  following the detection  of gravitational waves,  because of the versatility of massive graviton theories in resolving  multiple problems in cosmology and fundamental physics. In this work, we apply the method proposed in Rana et al.~\citep{Rana}, which consists of  looking for Yukawa-like fall off in the gravitational potential, to stacked galaxy cluster catalogs from three  disparate surveys. These include  catalogs from  2500 sq. degree SPT-SZ survey, the Planck all-sky SZ catalog, and a  redMaPPer selected catalog from  10,000 sq. degree of SDSS-DR8 data. The 90\% c.l. limits which we obtained  on the graviton mass using SPT, Planck and SDSS are:  $m_g < 4.73 \times 10^{-30}$ eV, $3.0 \times 10^{-30}$ eV, and $1.27 \times 10^{-30}$ eV respectively; or in terms of Compton wavelength are $\lambda_g >2.62 \times 10^{20}$ km, $4.12 \times 10^{20}$ km, $9.76 \times 10^{20}$ km. These limits are about five times more stringent than the previous best bound from galaxy clusters.
\end{abstract}
\maketitle

\section{Introduction}

There has been a renewed interest in obtaining
updated bounds on the graviton mass from solar system, galactic and  extragalactic  observations~\citep{Desai18,Rana,Will18,Zakharov18}, following the recent detection of gravitational waves from the LIGO detectors~\cite{LIGO1,LIGO2,LIGO3}. These massive gravity theories can solve multiple problems in cosmology and fundamental physics~\citep{Desai18}.
A comprehensive summary of all the  observational/experimental bounds on the mass of the graviton  as well as the sensitivity from future observations can be found in the comprehensive review  by Ref.~\cite{Derham16}, with a tabular summary  in Table 1 therein. These have also been succinctly summarized very recently  in Refs.~\cite{Desai18,Rana,Lee},
 wherein more details can be found. After these works, there have also been updated graviton mass bounds using solar system~\cite{Will18} and S2 star orbits~\cite{Zakharov18}. However, these are less stringent than the existing bounds documented in Ref.~\cite{Derham16}. Here,  for brevity, we only  focus on past and current limits on graviton mass using galaxy clusters, which look for consistency of the dynamics with a Yukawa potential.

Galaxy clusters are the biggest gravitationally  bound objects in the universe ~\cite{Vikhlininrev}. In recent years,  a large number of on-going (or recently completed) surveys  in the optical~\cite{DESc,SDSSc}, microwave~\cite{Bleem,ACT,Planck}, and X-ray~\cite{XCS}  have enabled the discovery of  large number of new galaxy clusters up to very high redshifts. These discoveries  have also enabled us to  constrain a large class of  modified theories of gravity~\cite{Rapetti,Raveri,Bocquet,Amendola,Vikhlinin,Lombriser,Terukina,Cataneo,Wilcox,Li,Silk,Rahvar,Ettori,Moffat13,Khoury17}. 
 
 However, despite the power  of galaxy clusters in testing beyond-GR theories, until this year there was only one published limit on graviton mass using galaxy clusters from 1974 by    Goldhaber and Nieto~\citep{Goldhaber74}. This  was obtained (following earlier arguments along the same lines by Hare~\citep{Hare}), by invoking Bertrand's theorem and  the fact that the characteristic size  of  Holmberg galaxy cluster catalog  is about 580~kpc~\cite{Holmberg}. They then obtain a limit of $m_g < 1.1 \times 10^{-29}$ eV, by looking for $\mathcal{O}(1)$ departures from Newtonian gravity at a distance of 580 kpc. However, as discussed in Refs.~\citep{Desai18,Rana}, the limit is extremely rough with no confidence level provided and the fundamental edifice used for obtaining this bound has recently been shown to be invalid~\citep{Mukherjee}.

In 2018, two different  works published updated limits on graviton mass using galaxy clusters. The first paper~\citep{Desai18} used  the dynamical mass modeling of Abell 1689 cluster~\citep{Hodson17,Nieu}, obtained using  high resolution weak and strong lensing data~\cite{Umetsu08,Umetsu15} to look for deviations between  the accelerations from Yukawa and Newtonian potentials.  From this analysis, a 90\% c.l. upper limit of $m_g < 1.37 \times 10^{-29}$ eV or $\lambda_g>9.1 \times 10^{19}$ km was obtained. Soon thereafter, Rana et al~\cite{Rana} (hereafter R18) obtained limits on graviton mass by looking for deviations between Newtonian and Yukawa acceleration profiles and stacking the galaxy cluster catalogs from the ACT SZ (Sunyaev-Zeldovich) survey~\citep{ACT} and also  using the weak lensing catalogs from the LoCuSS collaboration~\citep{Locuss}. From this analysis, they obtained a much more sensitive limit of $m_g <5.3 \times 10^{-30}$ eV, using the weak lensing dataset, and $m_g <8.3  \times 10^{-30}$ eV using the SZ dataset. Here, we apply the same formalism and methodology as in R18 to some of the biggest galaxy cluster catalogs in optical and SZ with well-calibrated masses, to constrain the graviton mass. 

This manuscript is organized as follows. We discuss the method used to constrain the graviton mass from stacked cluster catalogs in Section~\ref{sec:method}. The dataset used for this analysis is outlined in Section~\ref{sec:catalogs}. Our results can be found in Section~\ref{sec:results}. 
We conclude in Section~\ref{sec:conclusions}.

\section{Methodology}
\label{sec:method}

Here, we briefly recap the salient features of the method used in R18 to obtain bounds on graviton mass and also use their notation.

In Newtonian gravity, gravitational acceleration follows the inverse square law. So, for a given mass of a galaxy  cluster say, M$_\Delta$ within a radius R$_\Delta$, the Newtonian acceleration will be:
\begin{equation}
a_{n} = \frac{G M_{\Delta}}{R_{\Delta}^2}
\label{newton}
\end{equation}
where $G$ is the  Gravitational constant and R$_\Delta$ is the distance from the core of cluster at which the density of galaxy cluster becomes $\Delta$ times the critical density $\rho_{c}$  for an Einstein-DeSitter universe at that epoch. $\Delta$ is usually referred to as the over-density in Cosmology literature.  The critical density is given  by $\rho_c= \dfrac{3H^2(z)}{8\pi G}$, where $H(z)$ is the Hubble parameter at redshift $z$. 
The mass of the galaxy cluster can be evaluated from the  density of the galaxy cluster within a radial distance of R$_\Delta$~\citep{Rana}:
\begin{equation}
M_{\Delta}=  \Delta \times \rho_c  \times \frac{4\pi}{3} R_{\Delta}^3
\label{massdel}
\end{equation}

Now, to quantify  the differences between a Newtonian  and Yukawa potential, we need to determine the gravitation acceleration  in a  Yukawa potential. This can be obtained from the gradient of the Yukawa potential and can be written as~\cite{Will97}:
\begin{equation}
a_{y}=   \frac{GM_{\Delta}}{R_{\Delta}}  \exp(-R_{\Delta}/ \lambda_g) \left( \frac{1}{R_{\Delta}} + \frac{1}{\lambda_g} \right)
\label{yuk}
\end{equation}
For our analysis, we need to write down  the equations for both these accelerations in terms of observables and eliminate unknowns such as the galaxy cluster radius. Therefore, plugging R$_\Delta$ from Eq.~\ref{massdel} in terms of $M_{\Delta}$;  a$_n$ from Eq.~\ref{newton} and a$_y$  from Eq.~\ref{yuk} can be re-written as follows:
\begin{equation}
a_{n}(z,M_{\Delta})= (G M_{\Delta})^{1/3} \left(\frac{ H^2(z) \Delta}{2}\right)^{2/3}
\label{newton_zM}
\end{equation}

and

%\begin{equation}
%a_{y}(z,M_{\Delta},\lambda_g)= (G M_{\Delta})^{2/3} \left(\frac{ H^2(z) \Delta}{2}\right)^{1/3} \times \\ \exp \left[-  \frac{1}{\lambda_g}\left(\frac{2M_{\Delta}G}{H^2(z)\Delta}\right)^{1/3}\right] \left[\frac{1}{\lambda_g}+ \left( \frac{H^2(z) \Delta}{2 M_{\Delta}G}\right)^{1/3}\right]
%\end{equation}
%\begin{dgroup}
\begin{dmath}
a_{y}(z,M_{\Delta},\lambda_g) \hiderel{=} (G M_{\Delta})^{2/3} \left(\frac{ H^2(z) \Delta}{2}\right)^{1/3} \times \\ \exp  \left[ - \frac{1}{\lambda_g}\left(\frac{2M_{\Delta}G}{H^2(z) \Delta}\right)^{1/3}\right] \left[ \frac{1}{\lambda_g}+ \left( \frac{H^2(z) \Delta}{2 M_{\Delta}G}\right)^{1/3}\right]
\label{yuk_zM}
\end{dmath}
%\end{dgroup}

To determine the upper bound on the graviton mass, we need to calculate the $\chi^2$ profile  for which we need the accelerations from both the potentials. For this, we shall use the inferred M$_\Delta$ from the masses of galaxy clusters obtained from SZ/optical surveys corresponding to  a given radius overdensity.

\begin{figure}
\centering
\includegraphics[width=0.5\textwidth]{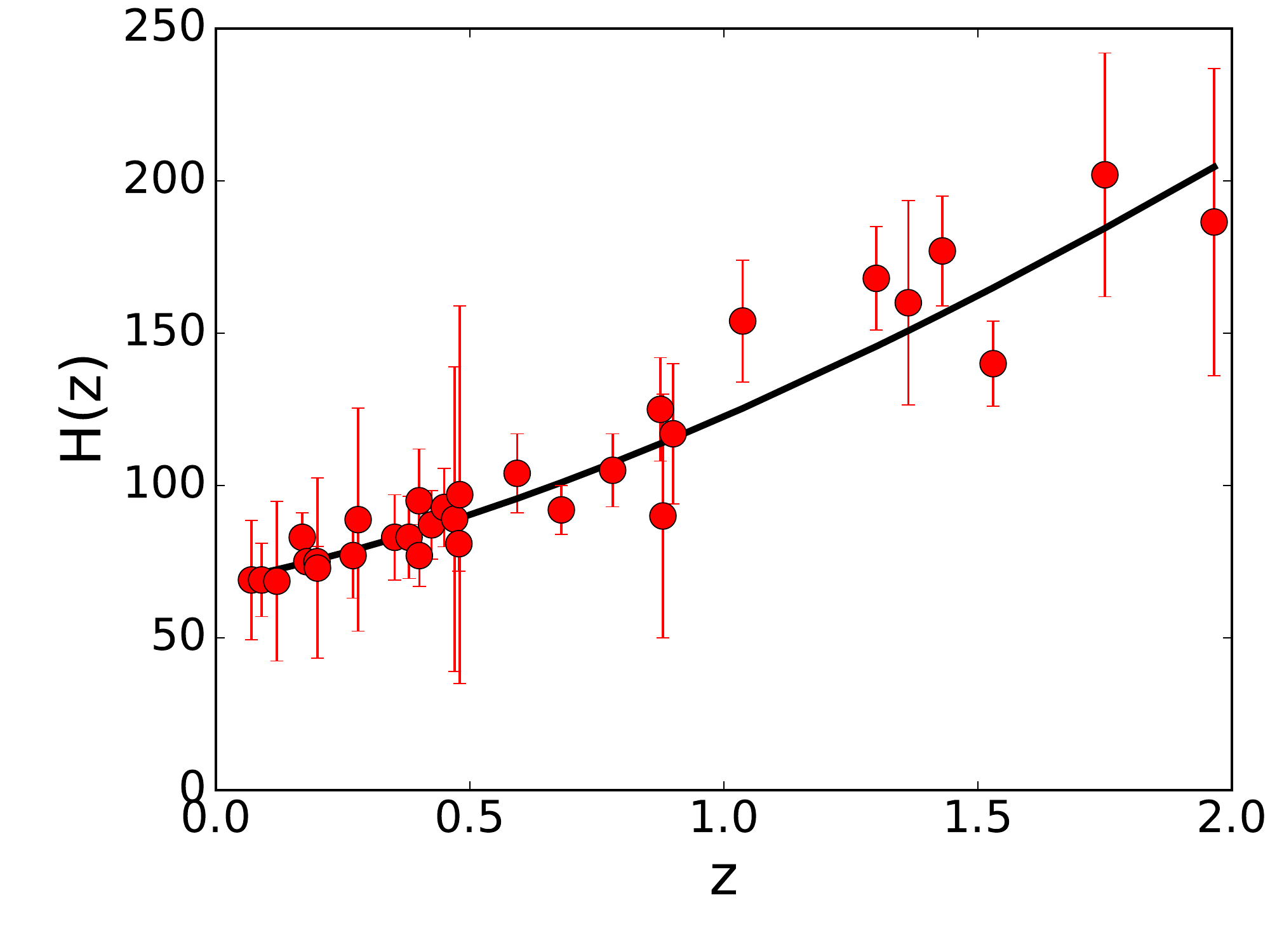}
\caption{$H(z)$ as a function of redshift, z. The red points along  with error bars denote the 31 measurements of H(z)~\cite{Ryan},  which are fitted against Eq.~\ref{H(z)} using least-squares fitting, giving us  $B=0.3 \pm 0.025$ and $C=0.65 \pm 0.078$.}
\label{fig:H(z)Vsz}
\end{figure}

In order to solve Eqs.~\ref{newton_zM} and ~\ref{yuk_zM}, we also need to evaluate the Hubble parameter $H(z)$ at any redshift $z$.
For this, we need independent sources of measurements of the Hubble parameter at different redshifts. We have used 31 $H(z)$ measurements obtained from  the cosmic chronometric technique within the redshift range of 0.07 $<$ z $<$ 1.965~\cite{Ryan}. To determine $H(z)$ value at any input redshift, we have fit the data to a non-linear 
function,  which mimics the behavior of H(z) in a flat $\Lambda$CDM cosmology~\cite{Ryan,Jesus}. The premise here is that a very tiny graviton mass does not largely change the expansion rate of the universe compared to $\Lambda$CDM.     
\begin{equation}
H(z) = A \sqrt{(B(1+z)^{3} + C)}.  
\label{H(z)}
\end{equation}
We have kept $A$ fixed at 70 km/sec/Mpc and the unknowns $B$ and $C$ in Eq.~\ref{H(z)} can be obtained using the least squares fitting technique as seen in Fig. \ref{fig:H(z)Vsz}. This parameterization allow us to evaluate the Hubble parameter  at any input redshift. Note that this procedure is slightly different from R18, who  have instead used the non-parametric procedure of Gaussian regression to model H(z). They also used H(z) measurements compiled in Ref.~\citep{Rana17}, instead of Ref.~\citep{Ryan}. 
Our best-fit values for $B$ and $C$ 
are given by  $B=0.3 \pm 0.025$ and $C=0.65 \pm 0.078$. Note that in $\Lambda$CDM cosmology for a flat universe, $A$, $B$ and $C$ are equal to $H_0$, $\Omega_M$ and $\Omega_{\Lambda}$ respectively. But for this analysis, they can be considered as arbitrary free parameters used  for evaluating H(z).

Once we have the mass for a given cluster, to quantify the deviations between Newtonian and Yukawa gravity, we construct a $\chi^2$ functional given by:
\begin{equation}
\chi^2= \sum\limits_{i=1}^N \left(\frac{a_{n}-a_{y}}{\sigma_a}\right)^2,
\label{eq:chi}
\end{equation}
where $a_n$ is the Newtonian acceleration defined in Eq.~\ref{newton_zM}; $a_y$ is the Yukawa acceleration defined in Eq.~\ref{yuk_zM}; and  $\sigma_a$ is the error in acceleration; $N$ is the total number of clusters and is defined in R18 as follows:
\begin{equation}
\sigma_a = \frac{a_n}{3}\sqrt{\left(\frac{\sigma_{M_{\Delta}}}{M_{\Delta}}\right)^2+ 16 \left(\frac{\sigma_H}{H(z)}\right)^2}
\end{equation}

As emphasized in R18, this procedure is not completely  model-independent. The mass estimates or mass calibration for the optical or SZ-selected galaxy cluster catalog is obtained using standard $\Lambda CDM$, whose edifice is based upon GR. Furthermore, the density profile used for their mass modeling is the well-known NFW profile~\citep{NFW}, which is obtained from $N$-body simulations  of particles interacting via Newtonian gravity. 
Another  premise here is that all the different components of the galaxy cluster (dark matter, gas, galaxies) follow the same functional form for the gravitational potential. If the dark matter candidate is a massive graviton~\cite{Postnov,Loeb,Amendola18}, then only the dark matter potential will show a Yukawa behavior. 
Furthermore, unlike in Ref.~\cite{Desai18}, we also do not make use of any direct or indirect measurements of the acceleration profile for any galaxy cluster. A completely ab-initio determination of the limit on graviton mass is beyond the scope of this work and will require extensive simulations and calibration of these simulations for a different graviton mass.
However, all current bounds on graviton mass involve some amount of model-dependence~\cite{Desai18}. 
\section{Cluster Catalogs}
\label{sec:catalogs}
Here, we  discuss  galaxy cluster catalogs used for our analyses. 
\subsection{SPT SZ}
The South Pole Telescope (SPT)~\cite{Carlstrom11} is a 10-meter telescope located at the South Pole and one of its main physics goals is to detect new galaxy clusters using the Sunyaev-Zeldovich (SZ)~\cite{SZ} effect. In 2011, SPT completed a 2,500 sq. deg survey of the southern skies
at 95, 150, and 220 GHz  and detected a total of 677 galaxy clusters with SNR $>$4.5~\cite{Bleem}. As of 2014, 516 of these were optically confirmed with measured redshifts and masses~\cite{Bleem}. Although, further optical/NIR follow-up and mass calibration  of all of these clusters  and cluster candidates is still ongoing using DES and other telescopes, for this analysis we use the measured masses and redshifts published in Bleem at al~\cite{Bleem}. The median mass of this sample is $M_{500} = 3.5 \times 10^{14} M_{\odot}$ and median redshift is $z_{med} = 0.55$.
The full SPT catalogs with masses and redshifts along with errors can be downloaded from
{\tt https://pole.uchicago.edu/public/data/\\sptsz-clusters/index.html}

\subsection{Planck SZ}
Planck was a satellite mission, operating from 2009 to 2013, with the microwave sky at nine different frequencies. In their third data release in 2015, Planck released an all-sky catalog of 1653 cluster candidates, of which 1203 were confirmed (as of 2015). The catalogs consisted of a union of three different matched-filter pipelines, as discussed in Ref.~\cite{Planck}.  Although optical follow-up of the remaining unconfirmed candidates is also ongoing (eg.~\cite{Planckszfollowup}), for this analysis we use the tabulated masses and redshifts from 907 clusters, distributed at the time of 2015 Planck data release. The mean mass of the confirmed clusters is  $M_{500} \sim 4.82 \times 10^{15} M_{\odot}$ and the mean redshift is 0.25. The 2015 Planck SZ catalog is available online at
{\tt https://heasarc.gsfc.nasa.gov/W3Browse/\\radio-catalog/plancksz.html}.

\subsection{SDSS DR8 redMaPPer}
redMaPPer is a red-sequence based galaxy cluster finding algorithm~\citep{SDSSc,DESc}, which has been applied to data from multiple photometric surveys such as SDSS, DES, HSC, CFHT, etc. For every galaxy cluster, redMaPPer provides a photometric redshift and an  optical  richness parameter called $\lambda$. For this analysis, we use the catalog of 26,111 clusters~\citep{SDSSc} obtained by running redMaPPer v6.3  on SDSS DR8~\citep{DR8} covering 10,000 sq. degrees. The calibration relation between $\lambda$ and the cluster mass  has been discussed in a number of papers (see the discussion in Ref~\cite{DESwl}). For this work, we use the richness-mass relation from Ref.~\cite{Simet},  which used weak lensing data for mass calibration of redMaPPer clusters as this accounts for both the statistical and systematic errors~\cite{DESwl}.
The redshift mass relation from this work is given by~\citep{Simet}:
\begin{equation}
M_{200} = 10^a \left(\frac{\lambda}{40}\right)^b h^{-1} M_{\odot}, 
\end{equation}
where $a=14.344 \pm 0.031$, and $b=1.33\pm 0.1$ and $h$ is the Hubble constant divided by 100 km/sec/Mpc. Note that the error in $a$ is the quadrature sum of the statistical and systematic errors documented in Ref.~\cite{Simet}, whereas the error in $b$ is the larger of the asymmetric (upper and lower) errors. The mean redshift and mass of this sample is at $z=0.36$ and $M_{200} =2.2 \times 10^{14} M_{\odot}  h^{-1}$. The SDSS redMaPPer v6.3 DR8 catalog can be downloaded from {\tt http://risa.stanford.edu/redmapper/v6.3/\\redmapper_dr8_public_v6.3_catalog.fits.gz}

\section{Results}
\begin{figure}
\centering
\includegraphics[width=0.5\textwidth]{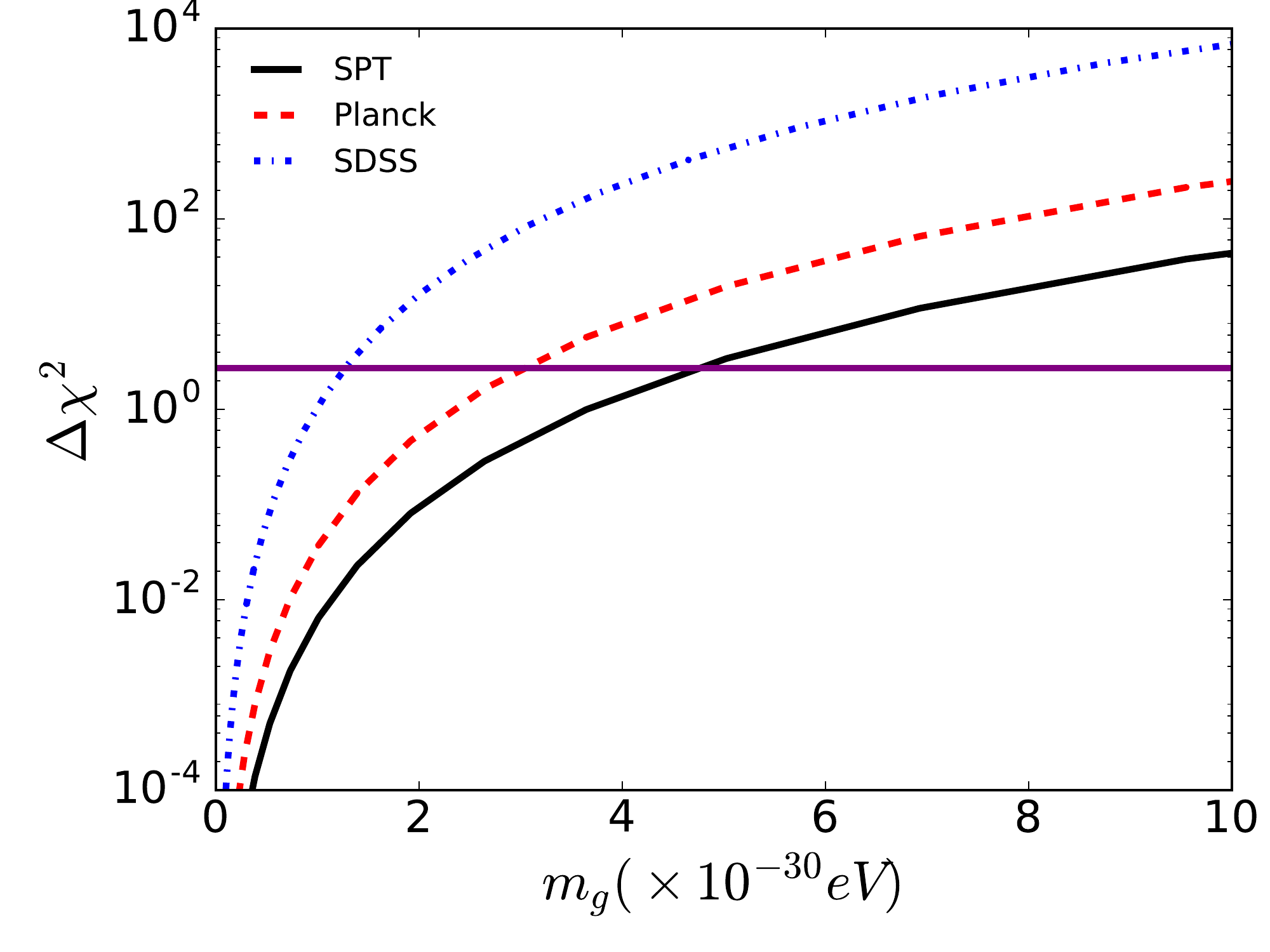}
\caption{$\Delta \chi^2$ as a function of graviton mass using stacked cluster catalogs from SPT, Planck and SDSS DR8 selected using redMaPPer. The solid magenta line at $\Delta \chi^2=2.71$ gives us the 90\% c.l. upper limit on the graviton mass. These upper limits correspond to   $m_g < 4.73 \times 10^{-30}$ eV, $3.0 \times 10^{-30}$ eV, and $1.27 \times 10^{-30}$ eV for SPT, Planck and SDSS respectively and are about five times more stringent than the corresponding limits in Ref.~\cite{Rana}.}
\label{fig:2}
\end{figure}

To calculate the upper limit on the graviton mass, we evaluated Eq.~\ref{eq:chi} separately for each of the three catalogs described in Sect.~\ref{sec:catalogs}. The 90\% c.l. upper limit on the graviton mass is obtained from the mass for which   $\Delta \chi^2=2.71$~\citep{NR}. As discussed in Ref.~\cite{Desai18},  despite the proximity to the physical boundary, the $\Delta \chi^2$ values needed for a given confidence interval remain the same as without  a physical boundary~\cite{Messier}. For this case, since $\chi^2_{min}$=0,  $\Delta \chi^2$ is trivially equal to the $\chi^2$ value evaluated in Eq.~\ref{eq:chi}. These trends of $\chi^2$ as a function of the graviton mass for all the three datasets are shown in Fig.~\ref{fig:2}. From these curves, the 90\% c.l. upper limits on the graviton mass from SPT, Planck and SDSS  are $m_g < 4.73 \times 10^{-30}$ eV, $3.0 \times 10^{-30}$ eV, and $1.27 \times 10^{-30}$ eV  respectively.
In terms of the Compton wavelength, the 90\% c.l. lower limits are given by $\lambda_g>  2.62 \times 10^{20}$ km for SPT, $\lambda_g>  4.12 \times 10^{20}$ km for Planck, and $\lambda_g>  9.76 \times 10^{20}$ km for SDSS. For a direct comparison  to R18 results, we also report the 68\% c.l. upper bounds, which are $3.45 \times 10^{-31}$ eV, $2.33 \times 10^{-31}$ eV, and $9.8 \times 10^{-31}$ eV for SPT, Planck and SDSS respectively. These limits have been tabulated in Table~\ref{my-label}.

Therefore, the most stringent limits are for the SDSS sample, because of the larger number of clusters (26,111).
Our limits are a factor of five more sensitive than those obtained  by R18 using the LoCuSS weak lensing  catalog and about an order of magnitude more sensitive then the limit using the  Abell 1689 cluster~\citep{Desai18}. However, this is  still two orders of magnitude less stringent then the weak lensing bound obtained in Ref.~\cite{Choudhury}, although there is some circularity used in obtaining  that limit~\citep{Desai18}.
\begin{table*}
		\centering
		\caption{Tabular summary of our 90\% c.l. (upper) limits  on the graviton mass (m$_g$) and (lower) limits  on the Compton wavelength ($\lambda_g$) for SPT, Planck and SDSS catalogs.}
		\label{my-label}
		\begin{tabular}{|l|l|l|l|l|}
			\hline
			%\multicolumn{5}{|l|}{\textbf{\begin{tabular}[c]{@{}l@{}}Greatest Lower Bound on graviton mass m$_g$ (in eV) and Lowest \\ Upper Bound on Compton Wavelength $\lambda_g$ (in Mpc)\end{tabular}}}                                                                                                                                                            \\ \hline
			\begin{tabular}[c]{@{}l@{}}Catalog\\ name\end{tabular} & \begin{tabular}[c]{@{}l@{}}No. \\ of clusters\end{tabular} & Type & m$_g < $ (eV)                                                                                             & $\lambda_g > $ (km)                                                                           \\  \hline 
			SPT                                                              & 516         &   SZ   &  4.73 $\times$ 10$^{-30}$   &  2.62 $\times$ 10$^{20}$                        \\ \hline
			Planck                                                           & 907         &   SZ   &  3.0 $\times$ 10$^{-30}$   &  4.12 $\times$ 10$^{20}$                        \\ \hline
			SDSS                                                             & 26111        &  Optical   &  1.27 $\times$ 10$^{-30}$   &  9.76 $\times$ 10$^{20}$                      \\ \hline
		\end{tabular}
\end{table*}

\label{sec:results}
\section{Conclusions}
\label{sec:conclusions}
Two recent papers from 2018~\cite{Desai18,Rana}  have obtained updated limits 
on graviton mass using galaxy clusters with complimentary techniques to supersede the extremely rough limit obtained 45  years ago~\citep{Goldhaber74}. Ref.~\citep{Desai18} looked for deviations between the estimated Yukawa acceleration  and observed acceleration for Abell 1689, using  the density profiles  for dark matter, gas and the central galaxy, to obtain a limit of $m_g<1.37\times 10^{-29}$ eV at 90\% c.l.)
The other paper~\cite{Rana} (R18) stacked galaxy clusters from two different surveys: one of them from the LoCuSS   weak lensing survey and the other from the ACT SZ survey. R18 looked for  deviations between Yukawa and Newtonian acceleration profiles and obtained a limit, which is about an order of magnitude   more stringent  than the one in Ref.~\citep{Desai18}, of   $m_g < 5.9 \times 10^{-30}$ eV.

We followed the same procedure as R18 and did more or less the same  analysis on three separate catalogs from recently completed galaxy cluster surveys in optical and microwave wavelength regimes. These include about 500 galaxy clusters from the 2500 sq. degree SPT-SZ survey~\citep{Bleem}, 900 galaxy clusters from the Planck all-sky SZ survey using their 2015 data release~\cite{Planck}, and a catalog of 26,000 galaxy clusters obtained from SDSS DR8 catalog with the redMaPPer algorithm. One minor difference between our analysis and R18, is that to obtain H(z), which is needed for the $\chi^2$ functional, instead of non-parametric smoothening, we fit the observational H(z) data to a function, which is close to its dependence on $z$ in a flat $\Lambda$CDM cosmology. Using these catalogs we obtain 90\% c.l. upper limit of $m_g < 4.73 \times 10^{-30}$ eV, $3.0 \times 10^{-30}$ eV, and $1.27 \times 10^{-30}$ eV from SPT, Planck, and SDSS  respectively (cf. Table~\ref{my-label}). These limits are about five times more stringent than the corresponding ones from R18.

Among the ongoing Stage-III dark energy experiments, the Dark Energy Survey is expected to discover 100,000 clusters covering 5,000 square degrees~\citep{DES} and the expected sensitivity using the same method is approximately $8 \times 10^{-31}$ eV. Stage IV dark energy experiments such as Euclid are expected to discover about $2 \times 10^6$ clusters,  of which one-fifth would be at $z \geq 1$~\citep{Sartoris}.
The estimated sensitivity to graviton mass using the Euclid sample will be about $4 \times 10^{-31}$ eV.
We caution however these expected limits  are back of the envelope estimates obtained by scaling the SDSS catalog. For a more detailed estimate, we need to use the mock catalogs obtained from the simulated skies of these surveys.
We note of course that one can also obtain such limits 
from single cluster analysis using the formalism in Ref.~\cite{Desai18}, which will be complementary to the techniques discussed in R18 and this work.

\begin{acknowledgements}
Sajal Gupta is supported by a DST-INSPIRE fellowship. 
\end{acknowledgements}
\bibliography{graviton2}
\end{document}